\documentclass[amssymb, amsmath, aps, prb, showpacs, nofootinbib,twocolumn]{revtex4} 
\usepackage{bm,url,graphicx,natbib}%

\long\def\symbolfootnote[#1]#2{\begingroup\def\thefootnote{\fnsymbol{footnote}}\footnote[#1]{#2}\endgroup}

\newcommand{\ETHz}{$E_{\rm{THz}}$\ }
\newcommand{\ETHzn}{$E_{\rm{THz}}$}
\newcommand{\InGaAsx}{$\rm{In_{1-x}Ga_{x}As}$}
\newcommand{\x}[2]{$\rm{In_{#1}Ga_{#2}As}$}
\newcommand{\tc}{$\tau_{\rm{c}}$\ }
\newcommand{\tcn}{$\tau_{\rm{c}}$}

\begin{document}
\title{Simulation and optimisation of terahertz emission from InGaAs and InP photoconductive switches}
\author{J.~Lloyd-Hughes}
\email{james.lloyd-hughes@physics.ox.ac.uk}

\author{E.~Castro-Camus}
\author{M.B.~Johnston}
 \email{m.johnston@physics.ox.ac.uk}
\affiliation{University of Oxford, Department of Physics, Clarendon Laboratory, Parks
Road, Oxford, OX1 3PU, United Kingdom}
\parbox{17.2cm}{\centering {\footnotesize SOLID STATE COMMUNICATIONS {\bf 136} 595
(2005)}}

\date{Uploaded to arxiv: 17th November 2005}

\begin{abstract}
{We simulate the terahertz emission from laterally-biased InGaAs and InP using
a three-dimensional carrier dynamics model in order to optimise the semiconductor material.
Incident pump-pulse parameters of current Ti:Sapphire and Er:fibre lasers are chosen, and the
simulation models the semiconductor's bandstructure using parabolic $\Gamma$, L and X valleys, and
heavy holes. The emitted terahertz radiation is propagated within the semiconductor and into free space
using a model based on the Drude-Lorentz dielectric function.
As the InGaAs alloy approaches InAs an increase in the emitted power is observed,
and this is attributed to a greater electron mobility.
Additionally, low-temperature grown and ion-implanted InGaAs are modelled using a finite
carrier trapping time. At sub-picosecond trapping times the terahertz bandwidth is found to increase
significantly at the cost of a reduced emission power.
}
\end{abstract}

 \maketitle

\fontsize{9}{10.8} \selectfont \pagestyle{myheadings} \markboth{}{\footnotesize{{\rm
LLOYD-HUGHES {\it et al.}, SOLID STATE COMMUNICATIONS {\bf 136} 595 (2005)}}}

\section{Introduction} \label{SEC:intro}
The ability to create and detect single-cycle pulses of coherent terahertz (THz)
radiation has enabled diverse studies in condensed matter physics. Recent examples
include a study of how surface plasmon polaritons propagate on semiconductor
gratings,\cite{prl93_256804} determining the low-energy vibrational modes of
oligomers,\cite{cpl03} and measuring the
time-dependent intersubband absorption in semiconductor quantum wells.\cite{prb70_155324}
Such studies have become possible using high power, broadband THz
radiation emitters based on the ultrafast separation of photoexcited carriers in
a photoconductive switch.\cite{jqe24_255} In this communication
we investigate how the choice of semiconductor alters THz emission, by employing a carrier
dynamics simulation to model InGaAs and InP photoconductive switches.

The use of a low-bandgap
semiconductor (such as {\InGaAsx} with an alloy fraction x\ $\leq0.47$) would enable cheaper,
more stable, turnkey THz spectroscopy setups based on erbium fibre (Er:fibre) lasers,
which can produce pulses as short as $65$\,fs at a central wavelength $\lambda=1.55\,\mu$m.\cite{oe11_594}
Combining a compact system of this type with communications wavelength optical fibres would enable THz
pulses to be generated and detected in small, extreme, or otherwise inaccessible
environments (e.g.\ in an endoscope or in the cavity of a pulsed magnet).

While GaAs-based photoconductive switch technology is well developed\cite{apl83_3117}, there are fewer
InP and InGaAs-based THz emitters described in the literature,\cite{ol18_1332,jjap43_7546,apl83_4113,apl86_051104,apl83_5551}.
Baker \emph{et.~al.} \cite{apl83_4113} reported THz emission up to 3\,THz from a low-temperature (LT) grown
\x{0.3}{0.7} photoconductive switch, using a Nd glass laser. Additionally, Suzuki and
Tonouchi\cite{apl86_051104} have measured the THz emission from photoconductive switches
made from \x{0.53}{0.47} implanted by 340\,keV, $1\times 10^{15}$\,cm$^{-2}$ Fe ions.
They used an Er:fibre laser with central wavelength 1.56\,$\mu$m and pulse duration
300\,fs, and measured THz power up to $\sim2$\,THz. Finally, Mangeney \emph{et.~al.} \cite{apl83_5551}
performed a similar experiment on \x{0.53}{0.47} implanted by Au$^+$ ions, with a carrier
lifetime of 0.9\,ps.

In order to investigate suitable semiconductors for photoconductive
switch THz emitters it is beneficial to adopt a modelling approach. The effect of
altering one parameter (or more) of the semiconductor material or laser can be investigated rapidly,
and an optimum combination sought. While analytical models of THz emission are relatively
straightforward to interpret, they often require phenomological assumptions (e.g.\ about
carrier mobilities).
In this paper we therefore employ a three-dimensional semi-classical carrier dynamics simulation
to model {\InGaAsx} and InP photoconductive switch THz emitters.


\section{Simulation details}
\label{SEC:sim}

The carrier dynamics model used herein is an extension of one used previously to model
the THz emission from InAs and GaAs surfaces under a magnetic field,\cite{prb02}
ion-implanted GaAs surfaces\cite{prb04} and GaAs photoconductive switches.\cite{prb05}
A set of $10^6$ pseudoparticles comprising extrinsic and
photogenerated carriers and fixed ions are used to simulate the semiconductor. At each
5\,fs step in time the model numerically finds the three-dimensional potential due to the
charge density, subject to the appropriate boundary conditions for the surface
and Schottky photoconductive switch contacts. Carrier-carrier, carrier-phonon and
carrier-impurity scattering mechanisms are included.\cite{prb02} The simulated particles are within a
box of size $x \times y \times z=6\times 6 \times 4\,\mu$m$^3$, subdivided into a grid of
$64\times64\times32$. The contacts of the
photoconductive switch are at $-3\leq x \leq -1\,\mu$m
and $1\leq x \leq 3\,\mu$m. A constant bias voltage of 5\,V between electrodes was used throughout this work, which
corresponds to typical experimental field strengths. The simulation starts at a time $t=-0.7$\,ps,
to allow the extrinsic carriers to equilibrate before the arrival of the incident pulse,
which has peak intensity at time $t=0.0$\,ps, and is centred at $x,y=0$.

Although the simulation assumes a parabolic band structure for simplicity, in the
modelling of THz emission from narrow band-gap semiconductor surfaces the inclusion
of the L-valley was found to create a more realistic distribution of carrier energies.\cite{prb02} In
photoconductive switches the average carrier energy is greater than for surface emitters, and intervalley (e.g.\
$\Gamma \rightleftharpoons L$) scattering by absorption or emission of an optical phonon
becomes more significant. Further, in the range $0.4 \lesssim x \lesssim 0.7$ the
X-valley energy gap of {\InGaAsx} falls below that of the L-valley. The simulation was therefore
extended to include the $X$ valley in addition to the $\Gamma$ and $L$ valleys.

\subsection{Properties of InGaAs} \label{SEC:InGaAs properties}

Values of semiconductor parameters for {\InGaAsx} at a particular alloy fraction $x$
were generated automatically within the simulation from the bowing parameter\cite{jap89_5815,NSM}.
When no bowing parameter was found in the literature, linear interpolation between InAs and GaAs was
used. The two-mode TO phonon energies for InGaAs of Ref.\ \cite{prb58_10452} were employed. For
consistency the intervalley deformation potentials were taken from Ref.\ \cite{jap68_1682},
the only reference found with values for GaAs, InAs and InP.

An inspection of the properties of InGaAs suggests potential advantages and disadvantages
as a THz emitter when compared with GaAs. Firstly, a lower $\Gamma$-valley effective
mass should result in a higher electron mobility, and greater THz power. Secondly, the intervalley deformation potential
D$_{\Gamma-L}$ is greater for InAs than GaAs\cite{jap68_1682}, resulting in a larger intervalley
scattering rate. When electrons scatter to the L or X valley they lose kinetic energy,
contributing to the deceleration of carriers that produces the negative peak in \ETHzn. The
lower absorption coefficient of InGaAs at $1.55\,\mu$m compared with GaAs at
800\,nm ($0.8, 1.2 \times 10^{6}$\,m$^{-1}$) is a disadvantage, because a lower $\alpha$ results in a lower bandwidth for
defect-laden semiconductors, owing to the photogeneration of fewer carriers within the damaged depth.
In addition, with a smaller $\alpha$ the emitted THz radiation will on average travel further through the
semiconductor (when the reflection geometry\cite{apl83_3117} is used), and will be
more significantly affected by absorption and dispersion.

Finally, since {\InGaAsx} exhibits two-mode TO phonon behaviour\cite{prb58_10452} the THz
emission spectra should exhibit strong absorption peaks near the TO phonon energies
(e.g.\ for x=0.47 the two modes have frequencies of 6.8 and 7.6\,THz). This will limit
the spectroscopic use of {\InGaAsx} photoconductive emitters in the frequency range near
and between the TO phonon modes of InAs and GaAs (6.4 and 8.1\,THz). The following
section describes how the two-mode TO phonon-polariton dispersion of InGaAs was included
in the simulation.

\subsection{THz propagation within photoconductive switch} \label{SEC:drude}
The radiated THz electric field \ETHz was calculated in the far-field approximation from the
simulated current density $\bf{J}$ using the relation
${\bf E}_{\rm{THz}} \propto ({1+\sqrt{\epsilon}})^{-1} \partial {\bf J}/\partial t$,\cite{shan04}
where the dielectric function $\epsilon$ was modelled using the well-known
Drude-Lorentz formula for TO phonon-polaritons,\cite{yu} which for two TO phonon modes is given by:
\begin{eqnarray}\label{EQN:TOphonon}
    \epsilon_{\rm{TO}}(\omega,\rm{x}) = \epsilon_{\infty}^{\rm{x}} &+&
        \frac{\epsilon_{s}^{\rm{x}}-\epsilon_{s}^{\rm{x}=1}}{1-\omega^2/\omega_{\rm{x}=0}^2-i \omega \Gamma/\omega_{\rm{x}=0}^2}
        \nonumber \\
      &+& \frac{\epsilon_{s}^{\rm{x}=1}-\epsilon_{\infty}^{\rm{x}}}{1-\omega^2/\omega_{\rm{x}=1}^2-i\omega \Gamma/\omega_{\rm{x}=1}^2}
\end{eqnarray}
\noindent where $\epsilon_s^{\rm{x}}$, $\epsilon_{\infty}^{\rm{x}}$ are the low and high
frequency dielectric constants at alloy fraction x. The angular frequency of the
GaAs-like (InAs-like) TO phonon is $\omega_{\rm{x}=1}$ ($\omega_{\rm{x}=0}$). The damping
rate was assumed to be $\Gamma=0.06$\,ps$^{-1}$ for both InAs-like and GaAs-like TO
phonon modes.\cite{yu} In order to estimate the thickness of semiconductor through which
the emitted THz must propagate we calculated the maximum THz electric field from each
`slice' parallel to the surface. Taking the mean of this distribution results in a
weighted propagation thickness of $\sim 0.5\,\mu$m. Assuming that the emitted THz
radiation is collected in the reflection geometry (to minimise absorption and dispersion
in the semiconductor material\cite{apl83_3117}) $0.5\,\mu$m is a reasonable estimate of
the propagation distance of THz radiation in an InGaAs photoconductive switch. We
therefore propagate \ETHz through this thickness of semiconductor, and into free space,
resulting in a characteristic reduction in spectral amplitude of emitted THz radiation
close to a TO phonon mode, and an enhancement just above the frequency of an LO phonon
(as seen in Figs.\ \ref{FIG:comparison}-\ref{FIG:varEgap}).

\subsection{Laser parameters} \label{SEC:lasers}
The simulation uses incident pulse parameters characteristic of current ultra-short
mode-locked lasers (i.e.\ a Gaussian spatial and temporal shape, and a transform-limited
Gaussian energy distribution). For the case of a 10\,fs Ti:sapphire laser (with central
wavelength $\lambda=800$\,nm, $\Delta \lambda = 80$\,nm, typical beam power
$P_{\rm{exp}}=400$\,mW and a repetition rate $R=75$\,MHz) we use a simulation power of
$10\,\mu$W and a Gaussian spot of standard deviation $\sigma_{x,y}=0.5\,\mu$m in order to
obtain the same photon flux as experimentally achievable. While amplified Er:fibre lasers
have been demonstrated with pulse durations of 65\,fs ($\lambda=1550$\,nm,
$\Delta \lambda > 100$\,nm, $P_{\rm{exp}}=110$\,mW, $R=67$\,MHz),\cite{oe11_594} sub-30\,fs pulses
widely tunable over the wavelength range 1130 to 1950\,nm may be generated by coupling
such Er:fibre laser light into a highly non-linear fibre.\cite{ol29_516} For excitation
around $1.55\,\mu$m we use $P=6\,\mu$W and $\sigma_{x,y}=0.5\,\mu$m, chosen to give the same
photon flux as for the Ti:sapphire, so that a comparison between \x{0.53}{0.47} switches
excited by $1.55\,\mu$m radiation, and GaAs excited by 800\,nm can be made in Section
\ref{SEC:results}. With the aim of providing a quick reference to the laser
parameters used we refer below to the 10\,fs, $\lambda=800$\,nm Ti:sapphire just described as
laser A, and the 65\,fs $\lambda=1.55\,\mu$m Er:fibre as laser B.

\section{Simulation results} \label{SEC:results}
\begin{figure}[tb]
    \centering
    \includegraphics[width=8cm]{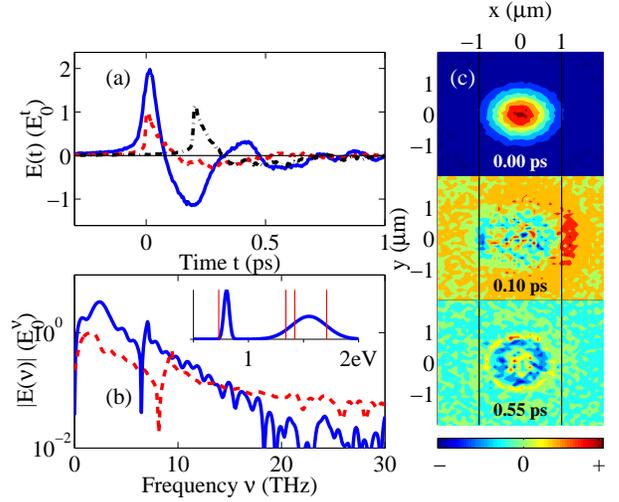}
    \caption{\label{FIG:comparison}(Colour online) (a) Simulated \ETHz for GaAs (dashed line) and InP (dash-dotted line, offset in time by +0.2\,ps for clarity)
    photoconductive switches excited by laser A, and InAs (solid line) excited by laser B. Both InP and InAs have a larger peak \ETHz than GaAs.
    (b) Fourier transform of \ETHz$(t)$ from (a), for InAs (solid) and GaAs (dashed line). The spectrum for
    InP (not shown) is substantially the same as that for GaAs (with its TO phonon absorption at 9.2\,THz).
    (Inset) Schematic energy distributions of a transform-limited 65\,fs pulse at $\lambda=$1.55\,$\mu$m (0.8\,eV)
    and a 10\,fs pulse at $\lambda=$800\,nm (1.55\,eV). Vertical lines at 0.73\,eV, 1.34\,eV, 1.42\,eV
    and 1.71\,eV respectively indicate $E_{\Gamma}$ for \x{0.53}{0.47}, InP and GaAs, and $E_{L}$ for GaAs.
    (c) Change in electron density between simulation time steps in the plane of the surface ($x-y$), averaged
    over the depth $z$ (see text for discussion). The black lines at $x=-1$\,$\mu$m and $x=1$\,$\mu$m mark
    the positions of the cathode and anode.
    }
\end{figure}

In Fig.\ \ref{FIG:comparison} we compare the simulated \ETHz for InP and GaAs photoconductive switches
excited by 10\,fs pulses from laser A with that obtained for InAs excited by 65\,fs pulses
from laser B. These parameters were chosen to enable a comparison between THz systems
based on current state-of-the-art lasers. In order to enable a comparison between the results presented
we normalize \ETHz in the time and frequency domains to the peak simulated values for a GaAs
photoconductive switch excited by laser A (with carrier trapping time \tc$=100$\,ps),
denoted $E_{0}^{t}$ and $E_{0}^{\nu}$ respectively. Similarly the
emitted THz power (defined as the square of $E_{\rm THz}(t)$ integrated over all times) is
normalized to $P_{0}$, the power for the GaAs emitter.

The peak value of \ETHz for InAs is larger than for GaAs, despite
the Er:fibre laser's power being lower than that of the Ti:sapphire (the photon flux is the same for both lasers).
This can be attributed in part to a higher fraction of incident photons having above-bandgap energy.
As the inset of Fig.\ \ref{FIG:comparison} indicates, almost $100\%$ of
photons from laser B have energy greater than $E_{\Gamma}\rm{(InAs)}=0.35$\,eV, while the
above-bandgap fraction for laser A is smaller ($\sim$80\%). The fact that {InAs} has a
lower effective mass in the $\Gamma$-valley than GaAs also contributes to the power
increase. Qualitatively, a smaller effective mass produces a larger mobility
$\mu=e\tau/m^*$, which should result in a larger transient current and therefore power.
However, this argument neglects changes in the carrier scattering time $\tau$,
and we return to this point in Section \ref{SEC:constDelta}.

Turning now to \ETHz for InP, we observe a slightly greater peak time-domain and spectral
amplitude than for GaAs, despite the larger $m_{\Gamma}^*$ (lower mobility) of InP. This is again related
to the energy distribution of the incident infrared pulses. Defining the excess energy of
the incident pulse as $\Delta=E_{\gamma}-E_{\Gamma}$ we see that
$\Delta=0.21$\,eV for InP, while $\Delta=0.13$\,eV for GaAs. A greater percentage of incident
photons are therefore absorbed in InP ($\sim$91\%) than GaAs (80\%), contributing to a larger
\ETHzn.

The simulated \ETHz for InAs exhibits an extra oscillatory component after
$t=0.30$\,ps, due to the onset of charge-density oscillations that can radiate THz radiation.
These plasma-type oscillations can be observed directly in the simulation as a change in electron density between
time steps (Fig.\ \ref{FIG:comparison}(c)). At $t=0.00$\,ps the rapid increase in electron density is due
to photoexcitation, and produces the first positive peak in $E_{\rm{THz}}$. By $t=0.10$\,ps there is an increased
electron density at the anode, while the first negative peak in \ETHz is at $t=0.2$\,ps (Fig.\ \ref{FIG:comparison}(a)),
and is primarily a result of the rapid momentum scattering of electrons. The subsequent change in \ETHz for $t>0.3$\,ps
can be identified as resulting from circular regions of altering
charge density (Fig.\ \ref{FIG:comparison}(c), $t=0.55$\,ps). Plasma oscillations have been observed
previously in the simulation of THz emission from InAs surfaces,\cite{prb02} and become less
noticeable as $\rm{x}\rightarrow 1$ due to GaAs's lower mobility.

\subsection{Carrier trapping time} \label{SEC:tauc}
Higher bandwidth terahertz emission can be achieved by introducing defects into a
semiconductor using a technique such as high energy ion implantation, or low-temperature
(LT) semiconductor growth. In both cases, carriers are trapped on sub-picosecond time
scales, with carrier trapping times \tc as short as 550\,fs in LT-\x{0.3}{0.7} (Ref.\
\cite{apl83_4113}) and 300\,fs in \x{0.53}{0.47} implanted with 2\,MeV Fe
ions.\cite{apl82_3913} When \tc becomes comparable with the duration of a single-cycle
THz pulse (\tc$\lesssim1$\,ps) bandwidth improvements can be observed. This effect may be
modelled using an exponential reduction in the photoexcited carrier density,\cite{prb04}
with a time constant \tcn.

\begin{figure}[tb]
    \centering
    \includegraphics[width=8cm]{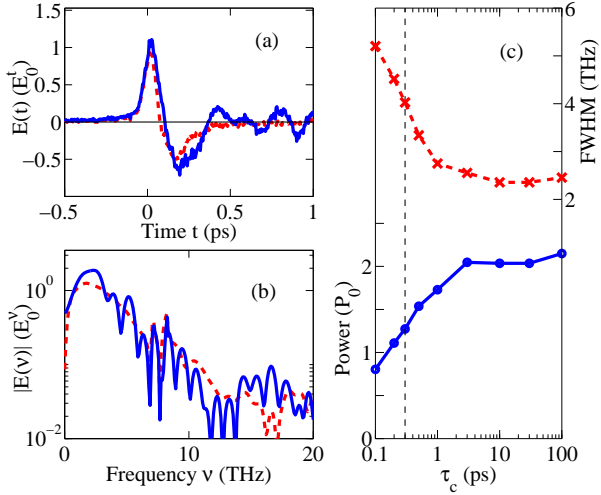}
    \caption{\label{FIG:tauc}(Colour online) (a) Simulated \ETHz for \x{0.53}{0.47} with
    \tc$=100$\,ps (solid line) and \tc$=0.3$\,ps (dashed line), using 65\,fs pulses with laser B's parameters.
    (b) Fourier transform of (a). In (c) the power (circles) and FWHM (crosses) are plotted for \tc$= 0.1-100$\,ps, showing
    a significant reduction in power and a bandwidth increase as \tc decreases below 1\,ps. The power is normalized
    to that of GaAs with \tc$=100$\,ps (excited by laser A), $P_0$. The dotted vertical line indicates \tc=0.3\,ps,
    used in subsequent figures.}
\end{figure}

In Fig.\ \ref{FIG:tauc} we present the simulated THz electric field \ETHz and calculated
spectra for \x{0.53}{0.47} excited by 65\,fs pulses of $1.55\,\mu$m centre wavelength
light, as a function of carrier trapping time. At shorter values of \tc the negative peak
in \ETHz arrives earlier, creating an increase in the spectral full-width-at-half-maximum (FWHM).
This trend can be seen in Fig.\ \ref{FIG:tauc}(c): when \tc$\lesssim1$\,ps there is a rapid increase in FWHM
at the cost of a reduced power, where the power is defined as $E^2_{\rm{THz}}(t)$ summed over all simulation times.
In the following we chose a value of \tc$=0.3$\,ps, achievable experimentally.\cite{apl82_3913}

\subsection{Varying alloy fraction at $\lambda=1.55\,\mu$m} \label{SEC:constLam}

\begin{figure}[tb]
    \centering
    \includegraphics[width=8cm]{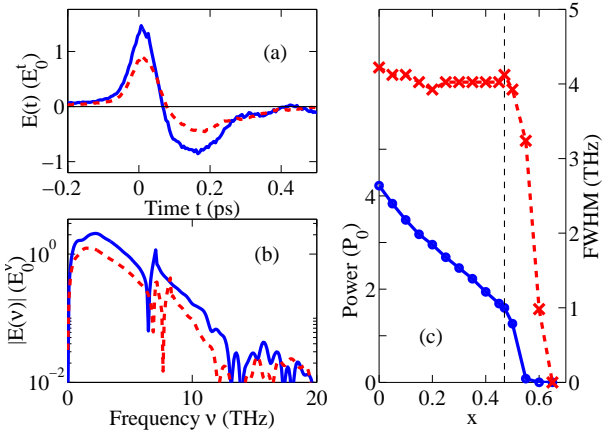}
    \caption{\label{FIG:varEgap}(Colour online) (a) and (b) Simulated \ETHz for InAs (solid lines) and
    \x{0.53}{0.47} (dashed lines) with \tc=0.3\,ps and excited by 65\,fs, $\lambda=1.55\mu$m pulses from
    laser B. (c) Power (circles)
    and FWHM (crosses) of simulated \ETHz as a function of x. The dotted vertical line indicates x=0.47.}
\end{figure}

With the intention of optimizing the semiconductor for excitation at
$1.55\,\mu$m we modelled the THz emission from {\InGaAsx} at alloy fractions x in
the range $0\leq \rm{x} \leq 0.7$. The
results of this set of simulations are shown in Fig.\ \ref{FIG:varEgap}. The emitted
power increases towards InAs to over four times larger than that of GaAs excited by
laser A, while the FWHM remains constant. When $\rm{x}\gtrsim 0.45$ both the power and FWHM
drop dramatically, since the excess incident pulse energy $\Delta$ becomes small and then
negative, and fewer photoexcited carriers are created.

While these results should be relatively straightforward to verify experimentally, the
density and kinetic energy of photoexcited carriers alter significantly when x is varied
and $\lambda$ kept constant. To try to assess better how the material choice alters THz emission
(independent of the laser's parameters) we performed a further set of simulations at
constant excess excitation energy $\Delta$, as reported in the next section.

\subsection{Varying alloy fraction at constant excess pulse energy $\Delta$} \label{SEC:constDelta}
Simulations were run over the full range of alloy fraction x with $\Delta=0.07$\,eV and
constant photoexcited carrier density. Pulsed Er:fibre lasers widely tunable in the range
1130 to 1950\,nm (1.1 to 0.64\,eV) have been recently developed,\cite{ol29_516} which
should allow experimental studies of this kind over a significant range of alloy fraction
($0.37 \lesssim \rm{x} \lesssim 0.77$).

\begin{figure}[tb]
    \centering
    \includegraphics[width=8cm]{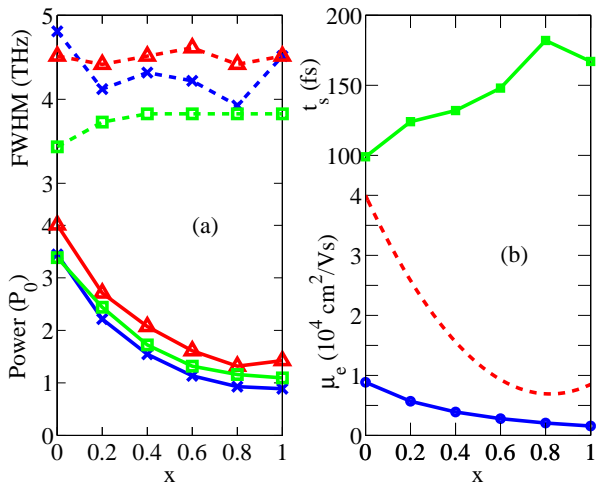}
    \caption{\label{FIG:varxvart}(Colour online) (a) Power (solid lines) and FWHM (dashed lines) of
    simulated THz electric field from {\InGaAsx} (\tc$= 0.3$\,ps), for incident pulse durations of 10\,fs (crosses),
    65\,fs (triangles), and 100\,fs (squares). The photon energy above the bandgap and the incident
    photon density were kept constant as the alloy fraction x was varied.
    (b) The average $\Gamma$ valley electron mobility (solid line) extracted from the simulation
    (see text for description), can be compared with the (low carrier energy) Hall mobility (dashed line).
    The time $t_s$ in which the applied electric field becomes screened (squares) increases towards GaAs.
}
\end{figure}

As Fig.\ \ref{FIG:varxvart}(a) shows, the emitted THz power increases
towards x=0, as found in Section \ref{SEC:constLam}. By extracting the total carrier momentum
scattering time $\tau$ from the simulation (summed over all scattering mechanisms, averaged over all electrons
in the $\Gamma$ valley, and taken at $t=0.00$\,ps)
we estimate the high-field electron mobility $\mu(\rm{x}) \simeq e\tau(\rm{x})/m_{\Gamma}^{*}(\rm{x})$. The relative
increase in $\mu(\rm{x})$ towards InAs is comparable with the enhancement in experimental low-field mobility
(Fig.\ \ref{FIG:varxvart}(b)), and results from a decrease in both the carrier-carrier and carrier-LO phonon scattering rates.

Shortly after the peak intensity of the incident pulse the electric field applied across a photoconductive switch is
completely screened by the build-up of the THz-emitting dipole. This occurs in a characteristic screening time
$t_s\sim100$\,fs for a semi-insulating GaAs photoconductive switch.\cite{prb05} By calculating the electric field
at the incident spot position ($x,y=0\,\mu$m) as
a function of time, it is possible to extract the screening time $t_s$, defined as the time at which the electric
field becomes zero. Figure \ref{FIG:varxvart}(b) indicates that $t_s$ increases as the Ga alloy fraction
$\rm{x}\rightarrow 1$, as a result of the lower electron mobility. It should be noted
that a shorter screening time does not necessarily mean a reduction in the THz power emitted: a higher mobility
can produce both a lower $t_s$ and a greater power.

This simulation set was repeated for incident pulse durations of 10\,fs and 100\,fs. In the 10\,fs case
the total power emitted (Fig.\ \ref{FIG:varxvart}(a)) is smaller than for 65\,fs across the entire
range of x, as fewer carriers are photoexcited. However, for 100\,fs the power is again lower
than the 65\,fs case, because when the later carriers are photoexcited the electric field is already partly screened.
How the THz power and bandwidth alter as a function of incident pulse duration was discussed for semi-insulating GaAs
in Ref.\ \cite{prb05}.
The FWHM of the 100\,fs spectra are lower than those of the 10\,fs and 65\,fs simulations, owing to the slow
rise in \ETHzn. Somewhat counterintuitively, the 10\,fs spectra do not have greater FWHM than the 65\,fs
over the whole range of x. This can be explained by noting that the FWHM
measures the width of the spectrum, and not the high-frequency spectral tail. Thus while
greater spectral amplitude is obtained at high frequencies for 10\,fs excitation
(e.g.\ Fig.\ \ref{FIG:comparison}(b)), the amplitude at low-frequency (which dominates the FWHM) is larger
at incident pulse durations of 65\,fs.

\section{Conclusion}\label{SEC:conclusion}
In summary, we have simulated the terahertz emission from {\InGaAsx} photoconductive
switches over the entire alloy fraction range $0 \leq \rm{x} \leq 1$. Two methods for increasing
the THz power from photoconductive switches were identified. Firstly, a larger electron
mobility resulted in a larger THz power as the Ga alloy fraction was decreased. This suggests that semiconductors
with high mobility/low effective mass may be used to improve THz power, for example
InAs ($m_{\Gamma}^*=0.022m_{e}$) or InSb ($m_{\Gamma}^*=0.014m_{e}$). Experimentally, a high resistivity
or semi-insulating InAs may be required to produce a photoconductive switch with a low dark current.
Secondly, increased THz power can be obtained by judicious choice of incident pulse energy, so that the majority of
photons are absorbed.

The spectra calculated from the simulation indicate that InAs photoconductive
switches have greater power than GaAs over the frequency range $0-15$\,THz. At higher
frequencies, the initial rise in \ETHz (dominated by the incident pulse duration) governs
the amplitude. This result seems to suggest that an InAs/65\,fs Er:fibre laser
setup should produce THz pulses that are at least as powerful as those from
10\,fs Ti:sapphire/GaAs setups.

\section{Acknowledgements}\label{SEC:ack}
The authors would like to thank the EPSRC, the Royal Society (UK)
and CONACyT (M$\acute{\rm{e}}$xico) for financial support.


\end{document}